\documentclass[structabstract]{aa}

\usepackage{graphicx}
\usepackage{lscape}
\usepackage{amsmath}
\usepackage{txfonts}
\usepackage{url}
\usepackage{natbib}
\usepackage{color}

\begin{document}
   \title{Solar system genealogy revealed by extinct short-lived radionuclides in meteorites}

   \author{Matthieu Gounelle\inst{1} and Georges Meynet\inst{2}}

   \authorrunning{Gounelle \& Meynet}

   \institute{Laboratoire de Min\'eralogie et de Cosmochimie du Mus\'eum, UMR 7202,
Mus\'eum National d'Histoire Naturelle \& CNRS, 75005 Paris, France
          \and Geneva Observatory, University of Geneva, Maillettes 51, CH-1290 Sauverny, Switzerland}

   \date{Received ; accepted }

 \abstract
   {Little is known {about} the stellar environment and the genealogy of our solar system. Short-lived radionuclides (SLRs, mean lifetime $\tau$ {shorter} than 100 Myr) that were present in the solar protoplanetary disk 4.56 Gyr ago could potentially provide insight into that key aspect of our history, were their origin understood.}
   {Previous models failed {to provide} a reasonable {explanation of} the abundance of two key SLRs, $^{26}$Al ($\tau_{26}$ = 1.1 Myr) and $^{60}$Fe ($\tau_{60}$ = 3.7 Myr), at the birth of the solar system
    by requiring unlikely astrophysical conditions.  Our aim is to propose a coherent and generic solution based on the most recent understanding of star-forming mechanisms.}
  {
  Iron-60 in the nascent solar system is shown to have been produced by a diversity of supernovae belonging to a first generation of stars in a {giant molecular cloud}.  Aluminum-26 is delivered into a dense collected shell by a single massive star wind belonging to a second star generation. The Sun formed in the collected shell as part of a third stellar generation. Aluminum-26 yields used in our calculation are based on
  new rotating stellar models in which $^{26}$Al is present in stellar winds during the star main sequence rather than during the Wolf-Rayet phase {alone}.   
  Our scenario eventually constrains the time sequence
  of the formation of the two stellar generations {that} just preceded
  the solar system formation, {along with} the number of stars born in these two generations.}
   {We propose a generic explanation for the past presence of SLRs in the nascent solar system, based on a  
 collect-injection-and-collapse mechanism, occurring on a diversity of spatial/temporal scales. In that model, the presence of SLRs with a diversity of mean lifetimes in the solar protoplanetary disk is simply the fossilized record of sequential star formation within a hierarchical interstellar medium (ISM). We identify the genealogy of our solar system's three star generations {earlier. }
   In particular, we show that our Sun was born together with a few hundred stars in a dense collected shell situated
   at a distance of 5-10 pc {from} a parent massive star having a mass {greater} than about 30 solar masses and belonging to a  cluster containing $\sim$ 1200 stars.}


   \maketitle
\section{Introduction}

Short-lived radionuclides (SLRs) are radioactive elements with mean lifetimes {under} 100 Myr {that} were incorporated {into} meteorites' primitive components such as calcium- {and} aluminum-rich inclusions (CAIs) or chondrules during the earliest evolution phases of our solar system. Understanding their origin has long been a major goal of cosmochemistry \citep{Russell2001}, as it is essential for constraining the stellar environment of the Sun at its birth \citep{Meyer2000}, as well as for establishing a chronology of the solar system{'s} first million years \citep{Keegan04}.

{The} SLRs with the longest mean lifetimes ($\gtrsim$ 5 Myr, such as $^{129}$I [$\tau_{129}$ = 23.5 Myr] or $^{244}$Pu [$\tau_{244}$ = 115 Myr]) have abundances compatible with that of the expected Galactic background {owing} to continuous star formation on kpc spatial scales and tens of Myr timescales \citep{Huss2009}. {Those} SLRs with shorter mean lifetimes ($\lesssim$ 5 Myr) appear to be in excess relative to that same background abundance. Among these, $^{10}$Be, $^{36}$Cl{,} and $^{41}$Ca can be made by solar energetic particle irradiation of the protoplanetary disk \citep{Gounelle2006,Duprat2007, Jacobsen2011}. For $^{10}$Be {alone}, {both} an interstellar origin \citep{Desch2004} {and} a solar-wind implantation model have also been evoked \citep{Bricker2010}. The origin of two key SLRs, $^{26}$Al ($\tau_{26}$ = 1.1 Myr) and $^{60}$Fe ($\tau_{60}$ = 3.7 Myr), remains elusive.

After decades of measurements within CAIs (the first solids to have formed in our protoplanetary disk), the solar system{'s} initial $^{26}$Al/$^{27}$Al ratio is well established at 5.3 $\times$ 10$^{-5}$ \citep{MacPherson1995,Jacobsen2008}, though its homogeneity is still subject to debate \citep{Villeneuve2009,Liu2012,Gounelle2005}. The situation is a bit more complicated for $^{60}$Fe {because} its record in CAIs is hampered by secondary processes {and} by nickel nucleosynthetic anomalies  \citep{Birck1988,Quitte2007}, whose origin is far from being understood   \citep{Steele2011}. Nickel-60 excesses attributed to the decay of  $^{60}$Fe were found in chondrules  \citep{Tachibana2003,Telus2011}, which are believed to have formed $\sim$1 Myr  after CAIs \citep{Villeneuve2009}. Based on these data, the presently accepted upper limit of $^{60}$Fe/$^{56}$Fe is $\sim$ 3 $\times$ 10$^{-7}$\citep{Dauphas2008, Gounelle2010, Telus2011}.

 To calculate the $^{60}$Fe and $^{26}$Al concentrations at the onset of solar system formation, in addition to the measured ratios presented above, we rely on the {\it protosolar} abundances given by Lodders (2003). With $^{56}$Fe/$^{1}$H = 3.2 $\times$ 10$^{-5}$, $^{27}$Al/$^{1}$H = 3.5 $\times$ 10$^{-6}$ and a hydrogen mass fraction of 0.71 \citep{Lodders2003}, we obtain the following concentrations in the nascent solar system for $^{26}$Al and $^{60}$Fe: C$_{\sun}$ [$^{26}$Al] = 3.3 $\times$ 10$^{-9}$ M$_{\sun}$/M$_{\sun}$ = 3.3 ppb (parts per billion) and C$_{\sun}$ [$^{60}$Fe] = 4.0 $\times$ 10$^{-10}$ M$_{\sun}$/M$_{\sun}$ = 0.4 ppb. The initial $^{26}$Al/$^{60}$Fe mass ratio was thus equal to 8.2. Such elevated concentrations of $^{26}$Al and $^{60}$Fe {need} to be explained, and are the subject or the present paper.

Though Asymptotic Giant Branch (AGB) stars have been proposed as a possible source of SLRs \citep{Wasserburg2006,Trigo2009,Lugaro2012}, massive stars (M $\ge$ 8 M$_{\sun}$) are the best candidates to account for {the presence of} $^{26}$Al and $^{60}$Fe in the nascent solar system. This is because massive stars at all stages of their evolution are present in star-forming regions, {unlike} AGB stars \citep{Kastner1994}.

The most massive stars (M $\gtrsim$ 25 M$_{\sun}$)  burn hydrogen for million{s of} years on the main sequence (MS) before they enter the short-lived Wolf-Rayet (WR) phase that precedes the supernova (SN) explosion. Massive stars lose their nucleosynthetic products to the Interstellar Medium (ISM) via strong winds (during the MS and WR phase) and during the SN explosion. Interestingly, while $^{60}$Fe is released only during the SN explosion, $^{26}$Al is released during the MS, the WR{,} and the SN phases \citep{Limongi2006, Palacios2005}.

In the classical SN model, first proposed by \citet{Cameron1977}, just after the discovery of $^{26}$Al \citep{Lee1976}, $^{26}$Al and $^{60}$Fe were delivered together by a single SN \citep{Boss2010, Ouellette2010} into the nearby solar protoplanetary disk or prestellar core. The distance, $r$,  at which {an} SN needs to be in order to inject {an} SLR at the solar abundance into a phase (prestellar core or protoplanetary disk) having a linear size $r_{0}$ reads {as}\citep{Cameron1995, Gounelle2008}:

$$
r= {r_0 \over 2} \sqrt{
{\eta_{\rm SN} Y_{\rm SN}\over M_{\rm SLR}}
e^{-\Delta/\tau},
} \eqno (1)
$$
where $\eta_{\rm SN}$ is the mixing efficiency of the SN ejecta with the receiving phase, $Y_{\rm SN}$ is the SN yield of the considered SLR, $M_{\rm SLR}$ is the solar system mass of the SLR, $\Delta$ the time elapsed between the release of the radioactive element by the source and its incorporation in the receiving phase{,} and $\tau$ the mean lifetime of the SLR under scrutiny.

Because the solar system abundance of  $^{26}$Al is far better constrained than that of $^{60}$Fe, we use the former SLR (C$_{\sun}$ [$^{26}$Al] = 3.3 ppb) to calculate the maximum distance at which the SN has to be from either the disk or the core to deliver SLRs at the solar abundance.
 Using the SN yields of massive stars {with} M $\le$ 60 M$_{\sun}$ (Huss et al. 2009) calculated by Woosley \& Heger (2007), we  obtain r $\le$ 0.4 pc for a disk of mass 0.013 M$_{\sun}$ \citep{Hayashi1985,Ouellette2005} and size r$_{0}$ = 100 AU with an injection efficiency of 0.7 \citep{Ouellette2010}, and r $\le$ 0.6 pc for a core of mass 2 M$_{\sun}$ and size r$_{0}$~=~0.058 pc with an injection efficiency of 0.02 \citep{Boss2010b}. The receiving phases parameters (mass and size) correspond to {the} observed properties of disks and cores adopted by the tenants of the single SN model \citep{Ouellette2005,Boss2010b}, while the injection efficiencies are estimated by the same authors \citep{Ouellette2010,Boss2010b}.  As  we conservatively assumed $\Delta$ = 0, the calculated distances are upper limits of the maximum distances. In other words, if we applied a decay interval of  $\Delta \gtrsim$ 1 Myr as required by all SN models (Huss et al. 2009), it would imply that the receiving
  phases (disk or core) lie at a few tenths of a parsec from the SN at most to receive $^{26}$Al at the solar abundance.

\begin{figure}
\includegraphics[width=0.38\textwidth]{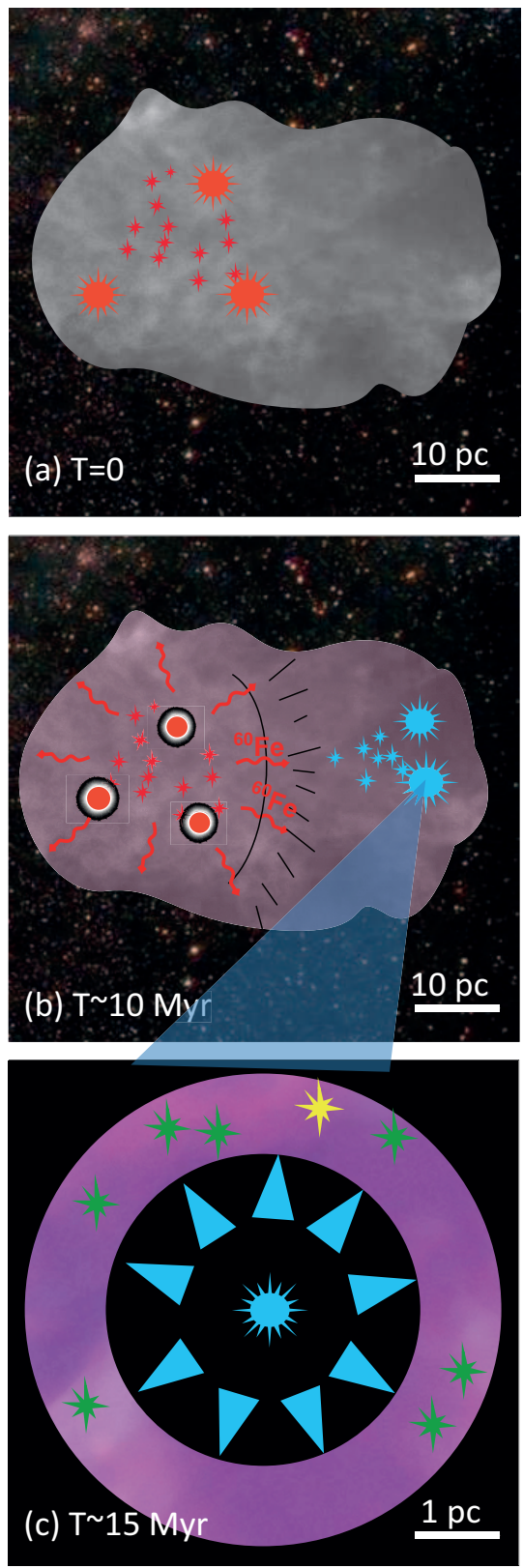}
   \caption{Sketch of the model described in the text (see Sect. 2). Star generations \#1, \#2, and \#3 are respectively in red, blue, and green. In panel (b), SNe remnants are shown with a dark contour. The reddish background symbolizes $^{60}$Fe delivered by the SNe belonging to the first star generation. In panel (c), the purple color of the shell symbolizes the combination of $^{26}$Al delivered by the single massive star wind from generation \#2 and the $^{60}$Fe delivered by generation \#1 stars. Our Sun, born in the circumstellar shell together with a few hundreds fellow low-mass stars (generation \#3), is shown in yellow.}
    \label{figsketch}
\end{figure}

It is very unlikely, if not impossible, {however,} to find a protoplanetary disk or a dense core that close to {an} SN. Before they explode as SNe, massive stars carve large ionized regions in the ISM (called HII regions) where the gas density is too low and temperature too high for star formation to take place \citep{Bally2008}. Observations show that even around a massive star that still needs to evolve for 2 Myr before it explodes as a SN, disks and cores are found several parsecs away \citep{Hartmann2005}, too far to receive $^{26}$Al and $^{60}$Fe at the solar abundance. In addition, because SNe ejecta are vastly enriched in $^{60}$Fe relative to $^{26}$Al and their respective solar abundances (Woosley \& Weaver 2007), all models relying on SN injection lead to a $^{26}$Al/$^{60}$Fe ratio {that is} far lower than the initial solar ratio of 8.2, unless special conditions are adopted \citep{Desch2011}.

Following the pioneering work of \citet{Arnould1997}, WR stars winds {have} recently { been} reconsidered as a specific source for $^{26}$Al {alone} \citep{Arnould2006, Gaidos2009, Tati2010}.
In the model of Gaidos et al. (2009), which considers injection of $^{26}$Al at the molecular cloud scale, very specific conditions (such as a precise timing between the formation of massive stars and the Sun or stellar clusters with an extremely large number of stars) are needed. In the stimulating model of \citet{Tati2010}, $^{26}$Al is delivered into a bow-shock-produced shell by a single runaway massive star moving with a velocity $\ge$ 20 km/s in a dense (n~$\sim$~100 cm$^{-3}$) star-forming region. 
At such a velocity ($\sim$ 20 pc/Myr), the runaway WR star considered by \citet{Tati2010} would escape any dense star-forming region with size $\ge$ 40 pc within 2 Myr, preventing the {collection} of dense gas well before the entry into the WR phase. In addition none of these models is generic (i.e. they fail to offer a common explanation for $^{60}$Fe and $^{26}$Al), nor do they constrain the solar system{'s} genealogy. Finally, because of the rarity of WR stars \citep{Crowther2007}, such models somehow require a special explanation for the formation of the solar system.

The goal of the present work is to identify a coherent model {that} accounts for the presence of $^{26}$Al and $^{60}$Fe in the nascent solar system and which is in line with the most recent astronomical observations of star-forming regions (Sect. 2). The proposed origin for $^{26}$Al (Sects. 2 and 3) is entirely original and relies on new rotating models of massive stars. The proposed explanation for $^{60}$Fe (Sect. 4) is an update of the Supernova Propagation \& Cloud Enrichment (SPACE) model elaborated by Gounelle et al. (2009). Combining these two results, a generic explanation is offered for the presence of SLRs having a diversity of mean lifetimes in the early solar system (Sect. 5). In Sect. 6, we discuss our results, focusing on the newly established solar system genealogy.

\section{Model sketch \label{SecPhymod}}

\begin{figure}
\includegraphics[width=0.52\textwidth]{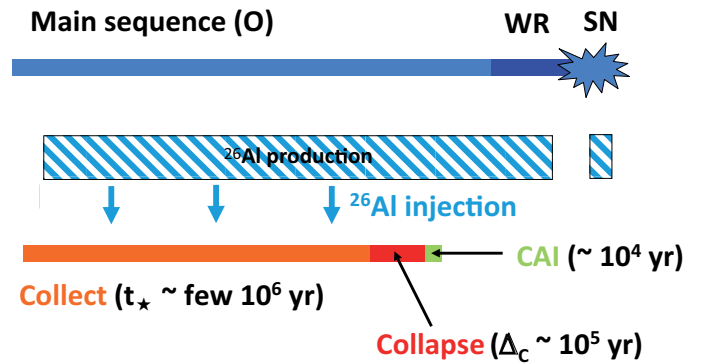}
   \caption{Timeline for $^{26}$Al injection in a dense collected shell around a massive star, starting approximately 10 Myr after the birth of stellar generation  \#1 and lasting $\sim$ t$_{\star}$. The upper part of the figure represents the time line of the central massive star of panel {\it c} in Fig. 1. In the middle part, the timeline of the  $^{26}$Al production and injection (arrows) is shown. Aluminum-26 is present in the wind some 10$^5$ yr after the star formation and absent when products of helium-burning appear at the surface (WC-type star) since $^{26}$Al
is destroyed in helium-burning cores. It is present while the star explodes as {an SN}. Aluminum-26 injection in the shell lasts during the whole collect{ion} phase and is assumed to stop during the collapse phase lasting a time $\Delta_C$ $\sim$ 10$^5$ yr. The lower part represents the timeline of the collected shell, which eventually collapses to produce a protoSun and a protoplanetary disk in which most CAIs form.}
    \label{fig_timeline}
\end{figure}

We propose to explain the SLR abundances in the nascent solar system within the framework of a common picture of star formation \citep{Hennebelle2009}, namely that of sequential star formation within a spatially and temporally structured {giant molecular cloud} (GMC).
We consider the following sequence of events (see  Fig.~\ref{figsketch}):
\begin{itemize}
\item Time 0 is that of the formation of a first generation  (\#1) of $N_1$ stars in a region \#1 of the GMC (see panel {\it a} of Fig.~\ref{figsketch}).
\item After a few Myr, massive stars from generation \#1 explode as {SNe} and start to
deliver $^{60}$Fe into a neighboring region \#2.
\item Five Myr after time 0 (see in Sect. 4.1 how this time is estimated), an $^{60}$Fe steady-state abundance (due to the balance between decay and production by SNe from generation \#1) is established in region \#2.
\item At t $\sim$10 Myr,  a second star generation (\#2, containing a total of $N_2$ stars) forms in region \#2, partly due to the compressive action of the generation \#1 SNe shockwaves onto the molecular cloud gas \citep{Preibisch2007}  (see panel {\it b} of Fig.~\ref{figsketch}).
\item From t  $\sim$ 10 Myr and for some Myr, the wind of one or two massive stars from generation \#2 will collect ISM gas to build a dense shell surrounding an HII region of radius 5-10 pc \citep{Deharveng2010}. That collected shell, which contains $^{60}$Fe originating {in} the SNe of the first generation, will be wind-enriched in $^{26}$Al via efficient turbulent mixing \citep{Koyama2002} during a time $t_{\star}$ lasting a few Myr (see Fig.~\ref{fig_timeline}).
\item At t $\sim$ 10 Myr + $t_{\star}$, aluminum-26 delivery ends when the dense shell fragments and collapses via a diversity of gravito-turbulent mechanisms{,} such as gravitational instabilities and ionization of a turbulent medium  \citep{Deharveng2010}{,} to form a third generation star cluster (\#3). The collapse phase lasts $\Delta_C$ $\sim$ 10$^5$ yr. Our Sun belongs to that third generation of stars (see panel {\it c} of Fig.~\ref{figsketch}). CAIs formed in the protoplanetary disk surrounding the protoSun {on} a timescale of a few 10$^4$ yr (Jacobsen et al. 2008) contains $^{26}$Al from the wind of the generation \#2 massive star and $^{60}$Fe from the generation \#1 SNe (Fig. \ref{fig_timeline}).

\end{itemize}

In the following, we show that such a scenario presents many realizations that permit a good fit of the quantities of $^{26}$Al and $^{60}$Fe at the solar system{'s} birth. In that respect, it provides
a more natural explanation than previous models{,} because it does not require special and/or unlikely conditions, and {leads to} understand{ing how} the solar system
form{s} {within} a broader perspective. The key point is that  $^{60}$Fe comes from the SNe of the first generation of stars, while $^{26}$Al comes from the wind of a {\it single} massive star belonging to a second star generation. This decoupling between the sources of $^{60}$Fe and of $^{26}$Al is essential to explain the observed $^{26}$Al/$^{60}$Fe ratio {when} the solar system form{ed}, and {it} does appear as a signature of the sequential star formation process.

\section{Origin of $^{26}$Al in the solar system}


\subsection{$^{26}$Al yields in massive stars winds}

\begin{figure}
\centering
\includegraphics[width=.52\textwidth]{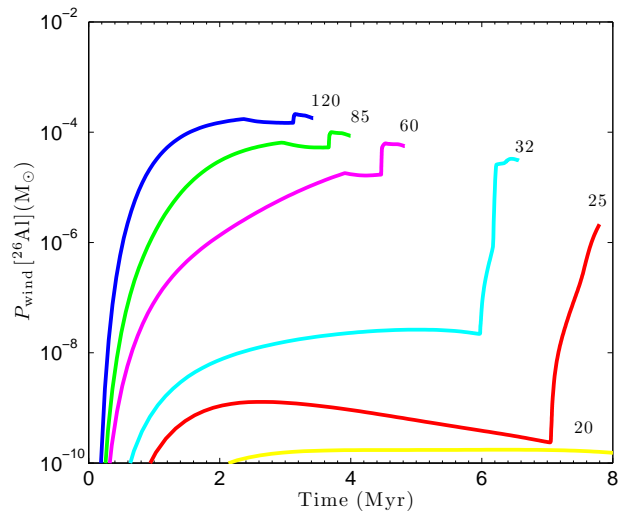}
   \caption{Integrated $^{26}$Al abundance produced by the stellar wind of massive stars as a function of time. Labels indicate star{s'} initial masses (in  M$_{\sun}$). These numbers take $^{26}$Al radioactive decay into account (see Eq. 1 of Arnould et al. 2006).}
    \label{figwind}
\end{figure}

The $^{26}$Al yields during the wind phase are calculated using new models for massive stars (Fig. \ref{figwind}). These models present two major improvements {over} previous ones \citep{Limongi2006,Palacios2005} considered so far by modelers \citep{Arnould1997, Gaidos2009, Tati2010}. First, they are built on newly determined initial solar abundances \citep{Asplund2009}.
	The solar abundances determined by \citet{Asplund2009} are similar to those shown by present-day massive stars in the solar neighborhood, and this composition is also quite similar to those expected in massive stars at the {formation} of the solar system \citep[see {T}able 5 in][]{Asplund2009}.

Second, these models account for the effects of axial rotation and improved mass loss rates \citep{Meynet2008}.
Rotational mixing is accounted for as in \citet{Palacios2005}. More precisely, rotational
mixing accounts for the effects of meridional currents and shear turbulence. The vertical shear diffusion was taken as given by \citet{Talon1997}, and the horizontal shear diffusion was taken as in \citet{Maeder2003}. A moderate overshooting is included. The radii of the convective cores given by the Schwarzschild's criterion are increased by 0.1 $H_p$, where $H_p$ is the pressure scale height estimated at the Schwarzschild boundary.
The mass loss rates are those of Vink et al. (2000, 2001) and of de Jager et al. (1988) outside
the application domain of the prescriptions by Vink et al. (2000, 2001). During the Wolf-Rayet phase, we use the mass loss
rates of Nugis \& Lamers (2000). The impact of rotation on the mass loss rates are accounted for
as in Maeder \& Meynet (2000).

These models reproduce reasonably well the observed number ratio in the solar vicinity of WR to O-type stars, of nitrogen-rich (WN stars) to carbon-rich (WC) WR stars{,} and of the transition WN/C to WR stars \citep{Meynet2008}. Because these ratios are sensitive to both rotation and mass loss, these good agreements provide some support to the way these processes are included in the present stellar models.

{We} stress here that rotational mixing allows some products synthesized in the core to appear at an early stage of the evolution of massive stars. For instance, while in non-rotating models, one has to wait until the H-rich envelope is nearly completely removed by the stellar winds in order for the star to eject $^{26}$Al, in rotating models the surface is already $^{26}$Al-enriched during earlier phases thanks to diffusion of $^{26}$Al from the core to the surface through the (shear unstable) radiative layers. This {lengthens} the period during which the wind is $^{26}$Al-enriched (Fig.  \ref{figwind}). This period is no longer reduced to the WR phase as in non-rotating models, but  also covers part of the MS phase.

\subsection{$^{26}$Al abundance in the collected shell}

\begin{figure}
\includegraphics[width=.52\textwidth]{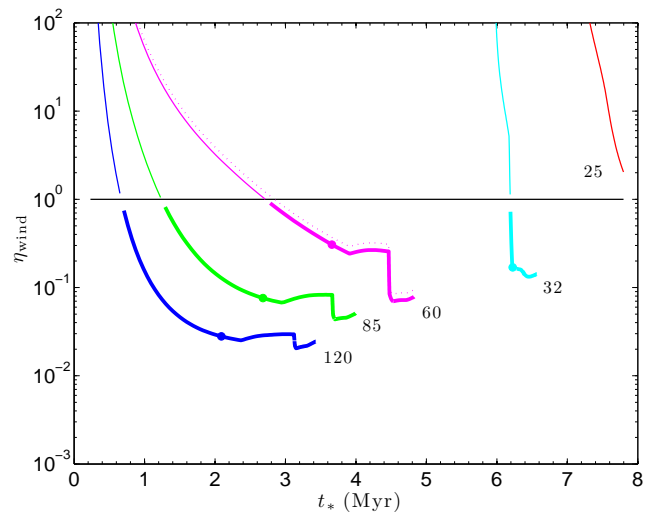}
   \caption{Thick lines show the acceptable model solutions, i.e. the values of $t_{\star}$ and $\eta_{\rm wind}$ for which the third generation stars have a solar $^{26}$Al abundance. By definition $\eta_{\rm wind}$ cannot be larger than 1. The filled circle shows for information the onset of the WR phase. The dotted line shows the 60 M$_{\sun}$ model with $\Delta_C$ = 0.5 Myr instead of 0.3 Myr (see text). }
    \label{fig_results}
\end{figure}

As already explained above,
the wind of one or two massive stars from generation \#2 (containing a total of $N_2$ stars) will collect ISM gas to build a dense shell. The dependence with time of aluminum-26 concentration in the shell, $C[^{26}{\rm Al}]$, reads {as} $C[^{26}{\rm Al}]= \eta_{\rm wind} P_{\rm wind}[^{26}{\rm Al}] (t) / M_{\rm shell} $, where $\eta_{\rm wind}$  is the mixing efficiency of the wind with the shell, $M_{\rm shell}$ is the mass of the collected shell and $P_{\rm wind}[^{26}{\rm Al}]$ is the quantity of non-decayed $^{26}$Al (i.e. survivor) present in the total wind ejecta of the massive star at time $t$. Imposing that the Sun forms in that shell after the collect and collapse phases, i.e. at a time $t_{\star}$ + $\Delta_C$ (Fig. \ref{fig_timeline}), we have

$$
C_{\sun}[^{26}{\rm Al}]= {\eta_{\rm wind} P_{\rm wind}[^{26}{\rm Al}] (t_{\star}) \over M_{\rm shell}}    e^{-\Delta_C/\tau_{26}}, \eqno (2)
$$
where $C_{\sun}[^{26}{\rm Al}]$ is the $^{26}$Al abundance at the onset of solar system formation.
The last term of the equation accounts for the decay of $^{26}$Al during a time
$\Delta_C$, which is the duration of the final collapse phase, when
the $^{26}$Al-carrying wind is conservatively assumed not to penetrate the shell because it has become too dense (Fig. \ref{fig_timeline}). 
 With $M_{\rm shell}$ = 1000 M$_{\sun}$ \citep{Zavagno2007}, $\Delta_C$ = 0.3 Myr \citep{Ward2007} and $C_{\sun}[^{26}{\rm Al}]$ = 3.3 ppb (see Sect. 1), {Eq.} (2) can be solved for $\eta_{\rm wind}$  and $t_{\star}$ (Fig. \ref{fig_results}).

As long as M $\ge$ 32 M$_{\sun}$, there is a {wide} diversity of solutions to our problem (Fig. \ref{fig_results}). An acceptable solution (among many others) is for example enrichment by a 85 M$_{\sun}$ star during t$_{\star}$~=~3 Myr with an injection efficiency $\eta_{\rm wind}$ = 0.07. The acceptable duration of enrichment (t$_{\star}$, see Fig. \ref{fig_timeline}) lasts from 0.65 to 6.2 Myr, while $\eta_{\rm wind}$ ranges from 2.1  $\times$ 10$^{-2}$ to 1. {For} many solutions, $^{26}$Al-enrichment ends before the onset of the WR phase; i.e., there {are many} values of t$_{\star}$ {that} are lower than t$_{\rm WR}$, where t$_{\rm WR}$ is the time of entry in the WR phase (see Fig.  \ref{fig_results}).

\subsection{Constraining generation \# 2: the parent cluster size}

\label{sec_clustersize}

\begin{figure}
\includegraphics[width=.52\textwidth]{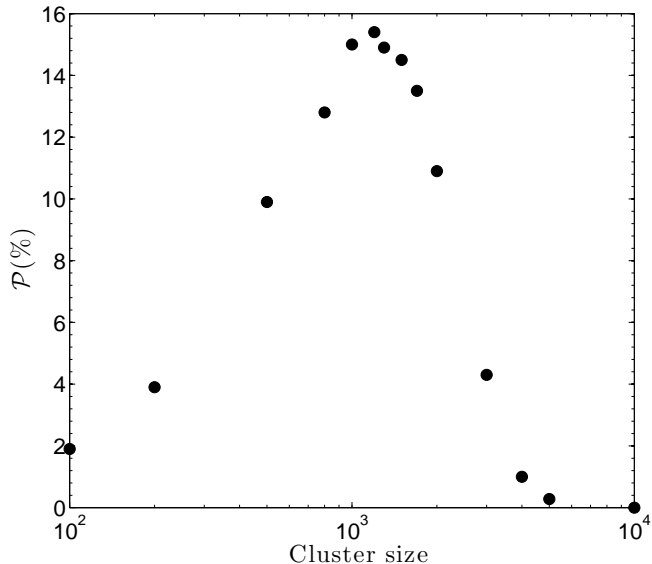}
   \caption{Probability $\cal P$ (in \%), as a function of the cluster size N, to realize the double condition: 1) the number of massive stars (M $\ge$ 8 M$_{\sun}$) is below n$_{\rm SB}$ = 5 and 2) one star at least has a mass {over} 32 M$_{\sun}$. For each point, the IMF (see text) was simulated 10000 times.}
    \label{figproba}
\end{figure}

{Unlike} other models, we can constrain the size of the parent cluster of our solar system. If cluster \#2 contains many massive stars, it is likely that a super wind bubble \citep{Parizot2004}  will form and that it will open into the ISM \citep{Gudel2008}, provoking {leakage} of $^{26}$Al atoms. Nice spherical shells are actually observed around single massive stars \citep{Deharveng2010} rather than around large clusters \citep{Gudel2008}. We can therefore constrain the size ($N_2$) of the cluster \#2 in imposing two requirements: 1) Cluster \#2 contains {fewer} than {five} stars more massive than 8 M$_{\sun}$ because superbubbles form when the number of massive stars is larger than  n$_{\rm SB}$ = 5 \citep{Higdon2005}. 2) Cluster \#2 contains at least one star more massive than 32 M$_{\sun}$, the minimum mass for which Eq. (2) has a solution (Fig. \ref{fig_results}). We call $\cal P$(N) the size-dependent probability distribution of a cluster containing N stars and satisfying conditions 1) and 2).

To calculate $\cal P$(N) we generate, for a given cluster size, 10000 realizations of the stellar initial mass function (IMF), and count the fraction of realizations which satisfy the two constraints exposed above. We use the IMF of \citet{Kroupa1993} mimicked by the generating function:

$$M = 0.08 + (0.19\xi^{1.55} + 0.05\xi^{0.6})/(1-\xi^{0.58}), \eqno (3)$$
where $\xi$ is a random number to be chosen between 0 and 1 \citep{Kroupa1993}. We consider only {those} distributions for which the most massive star is less massive than 150 M$_{\sun}$, a likely upper limit for stellar masses \citep{Weidner2006}. Using that generating function, the fraction f$_{\rm SN}$ of stars more massive than 8 M$_{\sun}$ (i.e. which will go SN) is 2.4 $\times$ 10$^{-3}$, and the average stellar mass is M$_\star$ = 0.5 M$_{\sun}$.

The calculated distribution probability is shown in Fig. \ref{figproba}. It is bell-shaped{,} and clusters with size $\sim$1200 stars have the highest probability ($\sim$15  \%) to satisfy the dual constraint evoked above. In~10  \% of the positive runs, clusters with 1200 stars contain two stars more massive than 32 M$_{\sun}$, which would both contribute to the $^{26}$Al enrichment. In such occurrences, the increased amount of $^{26}$Al would relax the (already loose) constraint on $\eta_{\rm wind}$.

\section{Origin of $^{60}$Fe in the solar system}
\subsection{Updating the SPACE model}

In our generalized model, $^{60}$Fe originates {in} SNe from the first generation of stars (see panel {\it a} in  Fig.1), as proposed in the Supernova Propagation And Cloud Enrichment (SPACE) model \citep{Gounelle2009}.
The calculations presented here are modified {when} taking the new (longer) mean lifetime of $^{60}$Fe \citep{Rugel2009} into account, and {when} adopting the SN yields of \citet{Woosley2007} rather than a combination of yields as in Gounelle et al. (2009).

{We} consider a set of $N_1$ stars coevally formed at time 0 and distributed in mass according to the initial mass function. The time-dependent
amount of $^{60}$Fe ejected by {SNe} in the interstellar medium, $P_{\rm SN}[^{60}{\rm Fe}](N_1,t)$, writes {as}

$$
P_{\rm SN}[^{60}{\rm Fe}](N_1,t)=\sum_{i=1}^{N_{\rm SN}}Y_{\rm SN_i}(^{60}{\rm Fe}) \; e^{-{{(t-t_i)}\over \tau_{60}}},\eqno (4)
$$
where $Y_{\rm SN_i}(^{60}{\rm Fe})$ are the $^{60}$Fe yields of the $i^{\rm th}$SN, $t_i$ is the explosion time of the $i^{\rm th}$ SN{,} $\tau_{60}$ is the $^{60}$Fe mean lifetime{, and} $N_{\rm SN}$ is the total number of SNe (i.e. stars more massive than 8 M$_{\sun}$) from generation \#1.The massive stars lifetimes are those calculated by  \citet{Schaller1992}.

 \begin{figure}
\includegraphics[width=.52\textwidth]{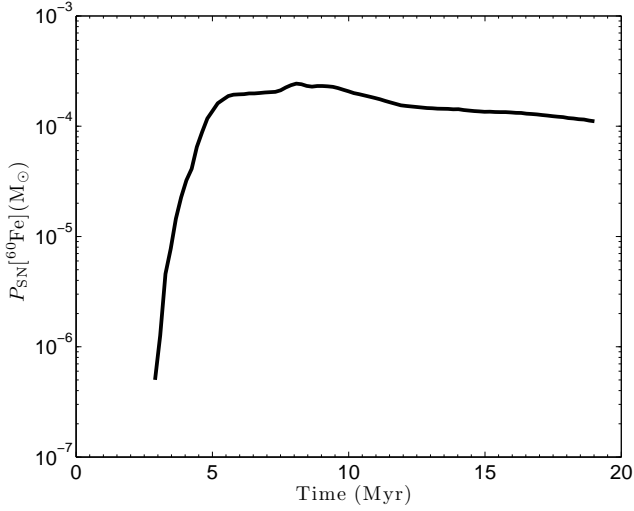}
   \caption{${P}_{\rm SN}$[$^{60}$Fe]($N_1$, t) for $N_1$ = 5000 stars, averaged over 1000 realizations of the IMF. The slight difference with Fig. 2 of Gounelle et al. (2009) is due to the longer mean lifetime of $^{60}$Fe.}
    \label{fig60Fe}
\end{figure}

To fix ideas, we adopt $N_1$ = 5000 stars and  perform the calculation 1000 times to take the stochastic nature of star formation into account. The evolution of  ${P}_{\rm SN}$[$^{60}$Fe](5000) with time is shown in Fig. \ref{fig60Fe}. It takes roughly 5 Myr to establish a steady state abundance of $^{60}$Fe (due to the balance of SN production and decay). We define the background (steady-state) value as $\hat{P}_{\rm SN}$[$^{60}$Fe]($N_1$) = ${P}_{\rm SN}$[$^{60}$Fe]($N_1$,
10 Myr), i.e. at the onset of star formation of generation \#2. In our fiducial case, $\hat{P}_{\rm SN}$[$^{60}$Fe](5000) = 2.1 $\times$ 10$^{-4}$ M$_{\sun}$ (Fig. \ref{fig60Fe}).
For each cluster size, 10000 realizations of the IMF were made to check the dependency of $\hat{P}_{\rm SN}$[$^{60}$Fe] with $N_1$. It is obviously linear and we find $\hat{P}_{\rm SN}[^{60}{\rm Fe}](N_1) = p_{\rm SN}[^{60}{\rm Fe}] \times N_1$,
with $p_{\rm SN}[^{60}{\rm Fe}]= 4.2 \times 10^{-8} {\rm M}_{\sun}$.

A GMC region having a mass $M_2$ receiving  $\hat{P}_{\rm SN}[^{60}{\rm Fe}]$ solar masses of $^{60}$Fe will have an $^{60}$Fe concentration of C[$^{60}$Fe] = $\gamma_{\rm SN}$ $\hat{P}_{\rm SN}[^{60}{\rm Fe}]$ / $M_2$ = $\gamma_{\rm SN}$  $\times$ $p_{\rm SN}$[$^{60}$Fe] $  N_1  / M_2$, where $\gamma_{\rm  SN}$ is the dilution factor of the ejecta in region \#2 and $N_2$ the number of stars formed in region \#2. With $M_2 = M_{\star}   N_2 / \epsilon$ where $\epsilon$  is the star formation efficiency and M$_\star$ is the average mass of a star (see Sect. 3.3), one obtains the simple relationship

$$C [^{60}{\rm Fe}] = \gamma_{\rm SN} \;  p_{\rm SN}[^{60}{\rm Fe}] \; N_1 \; {\epsilon \over N_2 \; M_\star}. \eqno (5)$$

With $p_{\rm SN}$[$^{60}$Fe]  = 4.2 $\times$ 10$^{-8}$, $\epsilon$ = 30  \% \citep{Elmegreen2007, Lada2003} and M$_\star$ = 0.5 M$_{\sun}$ (see Sect. \ref{sec_clustersize}), it gives the relationship
$$C[^{60}{\rm Fe}] = 25.2 \;   10^{-9} \;   \gamma_{\rm SN} \;   {N_1\over N_2} \; M_{\sun}  /  M_{\sun} \eqno (6)
.$$

Imposing that, at the onset of star formation in generation \#3, the $^{60}$Fe concentration established by the $N_1$ stars from the first generation of stars is equal to that of the nascent solar system ($C_{\sun}[^{60}{\rm Fe}]$), equation (6) reads {as}


$$C_{\sun}[^{60}{\rm Fe}] = 25.2 \; \gamma_{\rm SN}  \;  {N_1\over N_2}   \;  e^{-t/\tau_{60}}  \; {\rm ppb}, \eqno (7)$$
where $t$ is the time elapsed between the onset of star generation \#2 and of generation \#3 (that of our Sun).

\subsection{Constraining generation \# 1: the grandparent cluster size}

The quantity $t$ in Eq. (7), which is the duration between the onset of star generation \#2 and of generation \#3 can be written as the sum
of $t_{\star}$ and of $\Delta_C$.
With $N_2$ = 1200 stars (see Sect. 3.3) and 0.65 $\le t_{\star} \le$ 6.2 Myr (see Sect. 3.2),
we obtain 26 $\le \gamma_{\rm SN} \; N_{1} \le$ 108 or, to account for a possible revision of the initial solar system content of $^{60}$Fe, expressed as a function of the initial ratio $(^{60}{\rm Fe}/^{56}{\rm Fe})_0$:

$$26 \; \frac{(^{60}{\rm Fe}/^{56}{\rm Fe})_0}{3 \times 10^{-7}} \le \gamma_{\rm SN} \; N_{1} \le 108 \; \frac{(^{60}{\rm Fe}/^{56}{\rm Fe})_0}{3 \times 10^{-7}}. \eqno (8)$$

\begin{figure}
\includegraphics[width=0.52\textwidth]{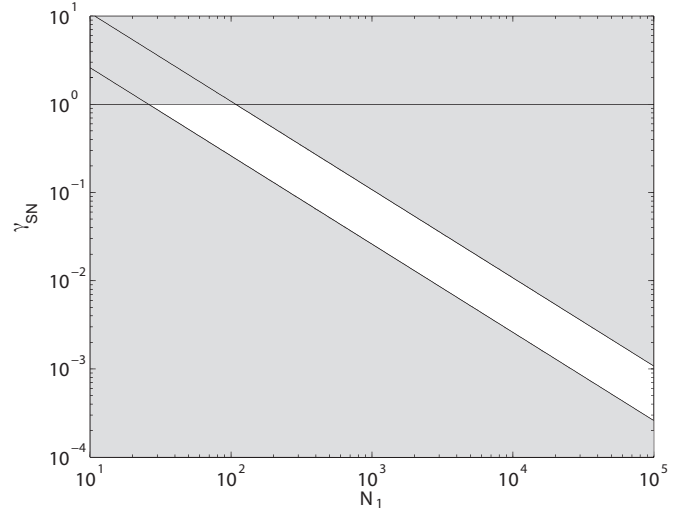}
   \caption{The dilution factor, $\gamma_{\rm SN} $ vs. $N_1$, the number of stars in star generation \#1. The allowed range of parameters is depicted in white and defined by $\gamma_{\rm SN} $ $\le$ 1 (by construction) and 26 $\le \gamma_{\rm SN} \;  N_{1} \le$ 108 (see Sect. 4.2).}
    \label{fig_etagamma}
\end{figure}

 The range of possible solutions ($N_1$, $ \gamma_{\rm SN}$) is shown in Fig. \ref{fig_etagamma}.
A first generation cluster with $N_1$ = 5000 stars and $ \gamma_{\rm SN}$ = 0.02 is a possible reasonable solution.
If, as recently suggested \citep{Moynier2011}, the solar system initial abundance of $^{60}$Fe was lower than the one usually accepted,  $ \gamma_{\rm SN}$ or $N_1$ could accordingly be decreased.

\section{The origin of SLRs: a generalized model}

It is therefore possible to identify the sequence of events that account for the presence of SLRs with a variety of mean lifetimes in the nascent solar system (Fig. 1).
\begin{itemize}
\item SLRs with the longest mean lifetimes ($\gtrsim$ 5 Myr) {and with} abundance compatible with a Galactic background origin, {come} from a large diversity of stars {on} the (kpc) scale of star complexes on a few 10s of Myr timescales \citep{Huss2009}.
\item $^{60}$Fe in the solar system was synthesized by a handful of SNe {on} the GMC (10s of pc) scale on a 5-10 Myr timescale.
\item $^{26}$Al was {finally} delivered by the wind of a single massive (M $\gtrsim$ 32 M$_{\sun}$) star on a few Myr timescale and on a 5-10 pc spatial scale.
\end{itemize}

The variability of temporal scales recorded by SLR{s'} mean lifetimes corresponds to the diversity of spatial scales at which star formation is observed \citep{Elmegreen2007}:
the {lon}ger the mean lifetime, the larger the scale of injection in the ISM. This spatio/temporal variability is simply the reflection of sequential star formation in a hierarchical ISM \citep{Elmegreen2007}. Sequential star formation is {an} important mechanism of star formation {that has been known for a long time} \citep{Elmegreen1977, Elmegreen2007, Hennebelle2009}. It is observed in a fossilized form in star-forming regions such as Orion \citep{Bally2008} or Scorpius-Centaurus \citep{Preibisch2007}, where OB subgroups with different ages are interpreted to have arisen from propagating star formation. It {has} also {been} caught in the act at the borders of HII regions \citep{Deharveng2010, Snider2009}, as well as {on} larger scales in the Carina star-forming region \citep{Chen2007}.

Though occurring {on a wide range} of spatial scales, a similar mechanism accounts for the {origins of} $^{26}$Al and $^{60}$Fe. For both radionuclides, ISM gas is collected by the same agent {as the one that} carries the SLR (SNe ejecta or massive star wind). After it has become dense enough, the collected gas is unstable to gravity and collapses to eventually form new stars. The SLRs {that} were contained in the collecting agent end up in the new generation of stars. In other words, given that the SNe shocks and massive star winds {that} contribute to the building of dense phases of the ISM also carry radioactive elements, it is not surprising that molecular clouds - and stars formed within them - are radioactive.

\section{Discussion}


		Our model therefore provides a natural explanation for the elusive presence of $^{26}$Al and $^{60}$Fe in the nascent solar system. It is generic {because}, both for $^{26}$Al and $^{60}$Fe, a collect-and-collapse mechanism \citep{Elmegreen1977} is at work, though operating {on} different scales. It does not suffer from the other difficulties encountered by the previous models. Because the $^{60}$Fe and $^{26}$Al have a decoupled origin (SN and wind, respectively), our model is not {faced} with the problem of the low $^{26}$Al/$^{60}$Fe ratio met by the single SN models \citep{Ouellette2005,Boss2010}. In addition, when the massive star goes SN, the disk is far {enough} away (5-10 pc) to avoid {the} disruption caused by the SN shockwave \citep{Chevalier2000} or over-injection of $^{60}$Fe{,} unlike the model of \citet{Gaidos2009}. More importantly, it {agrees} with the astronomical observations of star-forming regions and the accepted astrophysical mechanisms of star formation.

Our model leaves room for some imperfect mixing to account for the injection of $^{26}$Al by a stellar wind ($\eta_{\rm wind}$ $\le$ 1), unlike previous models {that} assume perfect mixing \citep{Gaidos2009, Tati2010}. Though it would be desirable to perform numerical simulations to calculate dust and gas injection efficiencies from the first principles, the values required in our model (as {low} as 0.02) can be reached easily. In fact, it was shown that the injection efficiency, at least for SNe ejecta, was multiplied by a factor of 3 when density was decreased by a factor of 2 \citep{Boss2010}. By construction, the collected shell has a {lower} density than that of a collapsing core, for which injection efficiencies are as high as 0.02 \citep{Boss2010b}.

The model is quite robust to parameter changes. First of all, it does not pretend to give a unique solution but {rather} a range of possible solutions. This diversity of solutions is intellectually satisfying {since} it would be {fairly} vain to pretend {to} exactly {identify} the star(s) responsible {for} the solar system $^{26}$Al and $^{60}$Fe enrichment relative to the Galactic background. It also means that there is ample room for parameter changes. Second, we have been quite conservative in the choice of our model inputs. This is exemplified by the adoption of new stellar models {with} metallicity lower than the one adopted by previous works. This lower metallicity implies lower initial content of $^{25}$Mg, which is the nucleus whose transformation produces $^{26}$Al during the core H-burning phase, and thus our models produce smaller quantities of $^{26}$Al  than the models of  \citet{Limongi2006}  and \citet{Palacios2005} used in previous work \citep{Arnould1997, Gaidos2009, Tati2010}. On the same line, we have conservatively considered that, during the final phase of collapse lasting $\Delta_C$ = 0.3 Myr, the shell was closed to $^{26}$Al injection, {unlike,} for example, the models of \citet{Boss2010}. Adopting an even {lon}ger isolation time would not change our conclusions (see the model with $\Delta_C$ = 0.5 Myr in Fig. \ref{fig_results}). The range of value{s} found for t$_{\star}$ is compatible with the observed timescales for the onset of star formation in a collected shell \citep{Deharveng2010}. {The} shell mass {agrees} with observations \citep{Zavagno2007,Deharveng2008} though it could be a factor of two higher \citep{Zavagno2007,Deharveng2008}. In any case, the assumed mass of the shell is 10$^5$ times {higher} than the disk of Ouellette et al. (2005), 500 times {more} than the core of Boss \& Keiser (2010), and {four} times {greater} than the most massive bow-shock shell of Tatischeff et al. (2010). It means that the probability of enrichment (not even taking the astrophysical context into account, Gounelle \& Meibom 2008), for a given massive star, is far {higher} in our collect-injection-and-collapse model than in any other model, considering that a single massive star is responsible for the solar system{'s} $^{26}$Al.

\begin{figure}
\includegraphics[width=0.52\textwidth]{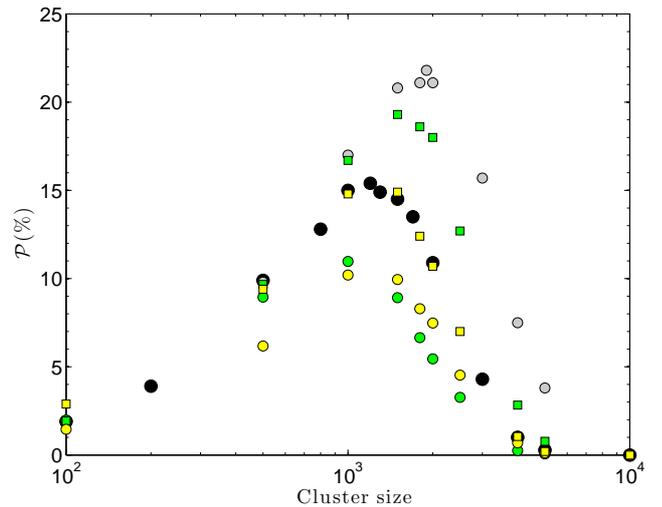}
   \caption{Probability $\cal P$ (in \%), as a function of the cluster size N, to realize the double condition: 1)  the number of stars with mass {higher} than M$_{\rm SN}$ is smaller than n$_{\rm SB}$ and 2) one star at least has a mass {greater} than M$_{\rm min}$, where the values n$_{\rm SB}$, M$_{\rm min}$, and M$_{\rm SN}$ are set to vary. For each point, the IMF (see text) was simulated 10000 times. Filled circles: same as Fig.  \ref{figproba}, i.e. n$_{\rm SB}$ = 5,  M$_{\rm min}$ = 32 M$_{\sun}$ and M$_{\rm SN}$ = 8 M$_{\sun}$. Gray circles: all being the same but M$_{\rm SN}$ = 10 M$_{\sun}$. Green circles: all being the same but n$_{\rm SB}$ = 4.  Green squares: all the same but n$_{\rm SB}$ = 6. Yellow circles: all  the same but M$_{\rm min}$ = 40 M$_{\sun}$. Yellow squares: all  the same but M$_{\rm min}$ = 25 M$_{\sun}$.      }
    \label{fig_clochevar}
\end{figure}

On a slightly different note, our reasoning does not critically depend on {adopting} the \citet{Woosley2007} {SN} yields. These were chosen because they are based on the solar abundances of \citet{Lodders2003}, which are almost identical to the newly determined solar abundances of \citet{Asplund2009}, unlike the older ones of \citet{Anders1989} used by other modelers of {SN} yields, such as \citet{Limongi2006}. These new abundances, calculated using 3D hydrodynamical models of the solar atmosphere, are supported by a high degree of internal consistency and by agreement with values obtained in the solar
neighborhood, though some discrepancy with heliosismology data persists \citep{Asplund2009}. It is also important to note that, though uncertain, the \citet{Woosley2007} SN yields are probably correct within a factor of a few since the IMF integrated $^{26}$Al/$^{60}$Fe ratio of SNe is comparable (within a factor of a few) to the one observed in the ISM by the INTEGRAL satellite \citep{Wang2007}. Finally,
adopting different yields (such as those of Limongi \& Chieffi 2006) would change the results by a factor of a few, not critical in the reasonings of {Sects.} 1 or  4, especially given the debates concerning the initial abundance of $^{60}$Fe (see below). 

An important aspect of our model is that it constrains the solar system genealogy; i.e., it provides estimates of the number of stars contained in the parent and grandparent clusters ($N_2$ and $N_1$, respectively). These cluster sizes are not constrained in the same fashion. In either {of these}cases, it is important to {keep} in mind that we  onlyhave access to an order of magnitude, as expected for a stellar nursery {that} was active 4.5 Gyr ago and whose only memory is preserved in meteorites' most primitive phases.

The best estimate of $N_2$ depends on a variety of numerical values {as} described in {Sect.} 3.3, namely the minimum number of stars, n$_{\rm SB}$, needed to initiate a super wind bubble expansion (identified to the minimum number of stars needed to create {an} SN-blown superbubble since the energy contained in stellar winds is comparable to {the energy} contained in SNe explosions, Parizot et al. 2004) and the minimum mass for which Eq. (2) has a solution. It also depends on the minimum mass for which a star goes {SN} which is M$_{\rm min}$ = 8 M$_\odot$  but
 is somehow model dependent \citep{Smartt2009}. To check that dependency, we explored the variations {in} $\cal P$(N) with these parameters (Fig. \ref{fig_clochevar}). Even allowing a relatively large and unlikely exploration of the parameter space, it is easy to see that the most probable value of $N_2$ always lies between 1000 and 2000. This exploration of parameters also encompasses  IMF formulations other than the one in Kroupa et al. (1993).


Given the dependency of the size of generation \#1 on the initial abundance of $^{60}$Fe and on $ \gamma_{\rm SN}$ (see Eq. (8)),
it is obvious that the value of $N_1$ is at present poorly known. It is not surprising that the size of the grandparent generation is less precisely constrained than the parent generation. As in any family, the further back in time one goes, the more difficult it is to retrieve precise information on ancestors. More knowledge will be gained on $N_1$ when analysts {have} converged on the initial solar system value of $^{60}$Fe \citep{Quitte2010,Moynier2011}{, along with} the development of numerical simulations {that} will help to better constrain the value of $ \gamma_{\rm SN}$.

The model can naturally account for some heterogeneity in the initial distribution of $^{26}$Al. Injection takes place on only one side of the dense shell. Because some time will be needed for $^{26}$Al carriers (gas or solid) to diffuse within the shell entire volume, it is expected that the first solids to form close to the protostar, as expected for CAIs \citep{Krot2009}, will not contain $^{26}$Al, as is observed in some meteoritic samples \citep{Liu2009, Makide2011, Weber1995}. Though the exact diffusion timescale is difficult to quantify without entering desired complex numerical simulations, it is expected to be {close to} the crossing time of the core, i.e. within the range of 0.5-1 Myr \citep{Bergin2007}, marginally compatible with the collapse timescale of a few 0.1 Myr \citep{Ward2007}.
Homogenization mechanisms will take place during the disk phase following the mechanisms presented by \citet{Boss2011}. Whether that homogenization goes to completion is at present unknown \citep{Villeneuve2009,Liu2012,Gounelle2005}. Given that $^{60}$Fe production {halts} some Myr and one star generation before the Sun{'s} formation, its distribution is expected to be homogeneous. At present, there is no experimental consensus on the homogeneous vs. heterogeneous distribution of $^{60}$Fe in the early solar system \citep{Dauphas2008,Quitte2010}.

Our model produces no undesired collateral enrichments \citep{Nichols1999,Gounelle2007}. Besides $^{26}$Al, massive stars winds do not contain other SLRs until the very last 3 $\times$10$^5$ yr of the star{'s} life \citep{Arnould2006}, during which $^{26}$Al injection is not considered. Given the high dilution factors of the wind needed to account for $^{26}$Al ($\le$ 10$^{-3}$), the oxygen isotopic composition of the shell is not modified. The shift in the oxygen isotopic composition is calculated to be less than a few per mil for all the considered models, which is far below our detection capability.

\section{Conclusion  and implications}

We have elucidated the origin of SLRs in the early solar system by developing a new model for $^{26}$Al {that} relies on a physical mechanism (collect + injection, collapse) similar in essence to the one presented recently for $^{60}$Fe \citep{Gounelle2009}. This new mechanism occurs naturally within a common mode of star formation, namely that of  sequential star formation within a {giant molecular cloud} \citep{Hennebelle2009}. Within a few milligrams of meteorites, SLRs therefore record physical mechanisms observed in the sky {on} scales varying from hundreds down to 1 pc.

The identified sequence of events establishes the genealogy of the solar system. Our Sun is the {great}-grandson of a star complex (generation 0) containing 10s of thousands of stars, the grandson of a large GMC core (generation 1) containing a few thousand stars, {and} the son of a massive ($\gtrsim$32 M$_{\sun}$) star belonging to a cluster (1000-2000 stars with a preferred value of 1200 stars, generation \#2) born later within the same GMC. Assuming 30  \% star formation efficiency \citep{Lada2003} and with an average stellar mass of M$_\star$ = 0.5 M$_{\sun}$, our Sun was born together with $\sim$ six hundred fellow stars in its natal 1000 M$_{\sun}$ shell.  Formation of the Sun in a relatively small clusteris in agreement with dynamical requirements, i.e. stability of planetary orbits, existence of the Kuiper belt object Sedna{,} and formation of the Oort cloud \citep{Adams2010, Brasser2011}.

Given the size of its cluster at birth (six hundred stars), it is not {in}conceivable that our Sun was coeval to a massive star. {Since t}he average number of massive stars in a cluster of size N {is} f$_{\rm SN}$ $\times$ N, with f$_{\rm SN}$ = 2.3 $\times$ 10$^{-3}$, clusters with six hundred stars might contain between 1 or 2 massive stars, most probably B stars, leaving  room for disk photo evaporation as suggested by \citet{Throop2005}.

It is expected that the few hundred stars born together with our Sun will share the same chemical and isotopic properties. These stars were true twins of our Sun. Given that stellar clusters dissipate on timescales of 100 Myr \citep{Adams2001}, that revolution timescales around the Galactic center are {on} the order of 200 Myr{,} and that stars can radially migrate over a few kpc in the Galaxy \citep{Roskar2008}, these fellow stars are now in totally unrelated places. Because the $^{26}$Al enrichment mechanism we have identified is generic, many other stars are, however, expected to contain $^{26}$Al at a level close to that of our solar system.

\begin{acknowledgements}
We sincerely thank an anonymous reviewer for {very} valuable comments {that} significantly helped to improve the paper. We are grateful to Thierry Montmerle for fruitful discussions, to S\'ebastien Fromang  for insights on dense core timescales as well as to Marc Chaussidon, Alessandro Morbidelli{,} and Edward Young for critical reading{s} of earlier versions of the manuscript. Figures were kindly produced by Michel Serrano. This work is partly supported by the Programme National de Plan\'etologie (PNP).
\end{acknowledgements}

\bibliographystyle{aa}
\bibliography{aa201219031RR_corrected_version_mg2}
\end{document}